\documentclass[11pt,twoside]{article}

\usepackage{asp2006}
\usepackage{epsf}
\usepackage{psfig}
\usepackage{lscape}
\usepackage{graphicx}

\begin{document}
\title{Digging into dark matter with weak gravitational lensing}   
\author{Richard Massey }   
\affil{Institute for Astronomy, Royal Observatory, Edinburgh EH9 3HJ, UK}    

\begin{abstract}
Ordinary baryonic particles (such as protons and neutrons) account for only one-sixth of the total
matter in the Universe. The remainder is a mysterious ``dark matter'' component, which does not
interact via the electromagnetic force and thus neither emits nor reflects light. However,
evidence is mounting for its gravitational influence. The past few years have
seen particular progress in observations of weak gravitational lensing, the slight
deflection of light from distant galaxies due to the curvature of space around
foreground mass. Recent surveys from the Hubble Space Telescope have provided direct proof
for dark matter, and the first measurements of its properties.
We review recent results, then
prospects and challenges for future gravitational lensing surveys.
\end{abstract}

\section{Introduction}

The concordance cosmological model poses a practical problem for observational astronomy.
Mounting evidence from many quarters now suggests that five sixths of the material in the
universe consists of exotic dark matter: outside the standard model of particle physics.
In particular, dark matter appears not to interact via the electromagnetic force, and
therefore neither emits, reflects nor absorbs light of any wavelength. Traditional modes
of astronomical observations are thus rendered blind.

On the other hand, dark matter is expected to interact via gravity.  It helps slow the
universe's expansion out of the big bang, and accumulates matter into the growth of
large-scale structure. Its gravitational influence is also its giveaway: even invisible
dark matter can be found and studied via its effect on visible particles. The most direct
technique, known as `gravitational lensing', studies the deflection of photons from a
distant light source, as they pass through gravitational fields (from galaxies, clusters
of gaalaxies or dark matter) along our line of sight. Gravitational light deflection is
similar in practice to the effect on light rays by an ordinary glass lens with a
refractive index different to that of air, magnifying and distorting an image. In
astrophysics, since a source cannot be viewed in the absence of lensing, any net
deflection of light is of little observational use. However, if light from two sides of a
resolved source are deflected by different amounts, its image appears measurably
distorted.

\begin{figure}[!t]
\begin{center}
\includegraphics[width = 200pt]{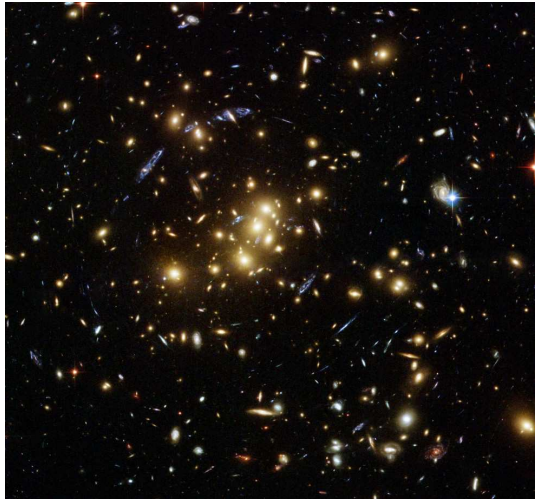}
\end{center}
\caption{\footnotesize{
Strong gravitational lensing around galaxy cluster  CL0024+17, demonstrating at least
three layers projected onto a single  2D image. The + shaped objects are nearby stars
in our own galaxy. 
The 
yellow, elliptical galaxies are members of the cluster, all at a similar redshift and
gravitationally bound. Also amongst this group  of galaxies is a halo of invisible dark
matter. The elongated blue objects are much more distant galaxies, unassociated with and
lying behind the cluster. Gravitational lensing has distorted their apparent images
into a series of blue tangential arcs. Figure credit: NASA/ESA/M.J. Jee (John Hopkins
University).}}\label{fig1}
\end{figure}

When such distortion is strong, a source no longer resemble astrophysical objects, and
observations of individual shapes can be used to infer the intervening gravitational
field. An example of `strong gravitational lensing' around a massive galaxy cluster is
shown in figure~\ref{fig1}. However, most lines of sight through the universe do not pass near
such a massive galaxy cluster. Farther from the dense cores of clusters, in the first
order limit of `weak gravitational lensing', the images of background galaxies appear
sheared: with their axis ratios typically changed by a few percent. Unfortunately,
individual galaxies have instrinsic shapes (elliptical bulges, spiral arms, etc.), which
cannot be disentangled from the small distortion. However, even galaxies on adjacent
lines of sight are usually far apart in 3D. Their true shapes must therefore be
uncorrelated, and the average shape must be a circle. Since the  lensing signal is
correlated over lines of sight separated by a few arcminutes, if sufficient galaxies 
galaxies can be imaged in that small patch of sky, their intrinsically circular mean
shape becomes observably elongated into an ellipse. Noise on this measurement is
dominated by the galaxies' intrinsic shapes, and about 100 typical galaxies are required
to bring the signal to noise to unity. This has only been achieved via the high
spatial resolution of the Hubble Space Telescope (HST).

\begin{figure}[!t]
\begin{center}
\includegraphics[scale = 0.2]{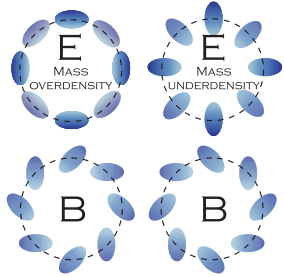}
\end{center}
\caption{The statistical signals sought by measurements of weak gravitational lensing are
slight but coherent distortions in the  shapes of distant galaxies. A tangential, circular
pattern is produced around a foreground mass overdensity, reminiscent of the  tangential arcs
of strong lensing. On much larger scales, an  opposite, radial pattern is
produced by foreground voids. Physical gravitational lensing produces  only these ``$E$-mode''
patterns. However, there is another degree of freedom in a shear (vector-like) field, and
spurious artefacts can typically mimic both. Measurements of ``$B$-mode'' patterns 
therefore provide a free test for residual systematic defects.}\label{fig2}
\end{figure}

To measure the distribution of foreground mass, patterns are sought in the apparent mean 
ellipticities of many thousands of distant galaxies. As illustrated in figure~\ref{fig2}, mass
overdensities in front of the galaxies produce a tangential $E$-mode pattern reminiscent
of the tangentially-aligned strong lensing arcs. Foreground mass underdensities or voids
produce radial $E$-mode patterns instead. Usefully, there is an additional degree of
freedom in the pattern of ellipticities. Curl-like $B$-modes are not produced by physical
gravitational lensing. However, noise and most instrumental systematics occupy both modes
equally. The measured $B$-mode signal can therefore act as an independent realisation of
noise and a warning of residual, uncorrected systematic effects in the desired
$E$-mode.

\section{Instrumental nuisances}

Although the weak lensing shear signal is coherent across several arcminutes, it is still
weak. Measuring the tiny distortions in the shapes of distant (therefore small and faint)
galaxies requires unusually precise control over imaging quality. For example, the
galaxies are inevitably viewed after convolution with the telescope's Point Spread
Function (PSF). For ground-based telescopes, this is dominated by turbulence in the
Earth's atmosphere, and is particularly troublesome. Even from space, diffraction through
the finite aperture of the primary mirror blurs the shapes of small galaxies in a manner
that can mimic or dilute a weak gravitational lensing signal. Furthermore, HST's
low-Earth orbit also brings it in and out of the shadow of the Earth. Thermal expansions
and contractions of only a few microns in its 13m length put it sufficiently out of focus
to alter the ellipticity of its PSF, and consequently that of galaxies, by an amount
comparable to the weak lensing signal. It is therefore necessary to model the shape of
the PSF from stars within each image. Unusually for extragalactic observations, the ideal
sky location for a weak lensing survey is therefore not necessarily at the galactic
poles.

The cameras currently aboard HST also suffer from a second instrumental problem. Harsh
radiation in the orbital environment has damaged the CCD detectors, creating charge traps
within the silicon lattice. At the end of each exposure, when photoelectrons are
transferred to readout electronics at the edge of the CCD, they can be temporarily
captured in these traps and released after a short delay. The main packet of
photoelectrons will typically have been moved several more pixels by the time the
captured electrons are released. The released electrons therefore produce a trail behind
each astronomical object. This mimics the weak lensing signal in sinister fashion. It
adds a coherent, spurious ellipticity to each object, and is a non-linear process that
affects faint and small galaxies more than bright, large ones. Worse still, because of
the particular orientation of the two CCDs in HST's Advanced Camera for Surveys, the
spurious ellipticity is confined to the $E$-mode signal. Current solutions include the
empirical calibration of extra ellipticity in the readout direction. Much
effort is also being invested in improved designs for future hardware and data reduction
software.

\section{Observations}

\subsection{Large-scale structure}

The Hubble Space Telescope COSMOS survey (Scoville et al. 2007) is almost as deep as the
Hubble Deep Field, and the largest optical survey ever obtained from space. At 2 square
degrees, and containing two million galaxies, it was designed to enlcose a contiguous
volume of the universe at redshift $z=1$ containing even the largest expected structures,
and at least one example of every type of environment. The single-colour HST imaging is
backed up by ground-based spectroscopy of ten thousand (and growing) galaxies, plus
multicolour imaging at about forty other wavelengths, from radio, through IR with
Spitzer, optical, UV with Galex and X-ray with both Chandra and XMM. As well as accurate
measurements of the shapes of galaxies, their (photometric) redshifts are also well
understood. The X-ray imaging is also sensitive to hot gas, typically found within dense
clusters of galaxies.

\begin{figure}[!ht]
\begin{center}
\includegraphics[scale = 0.2]{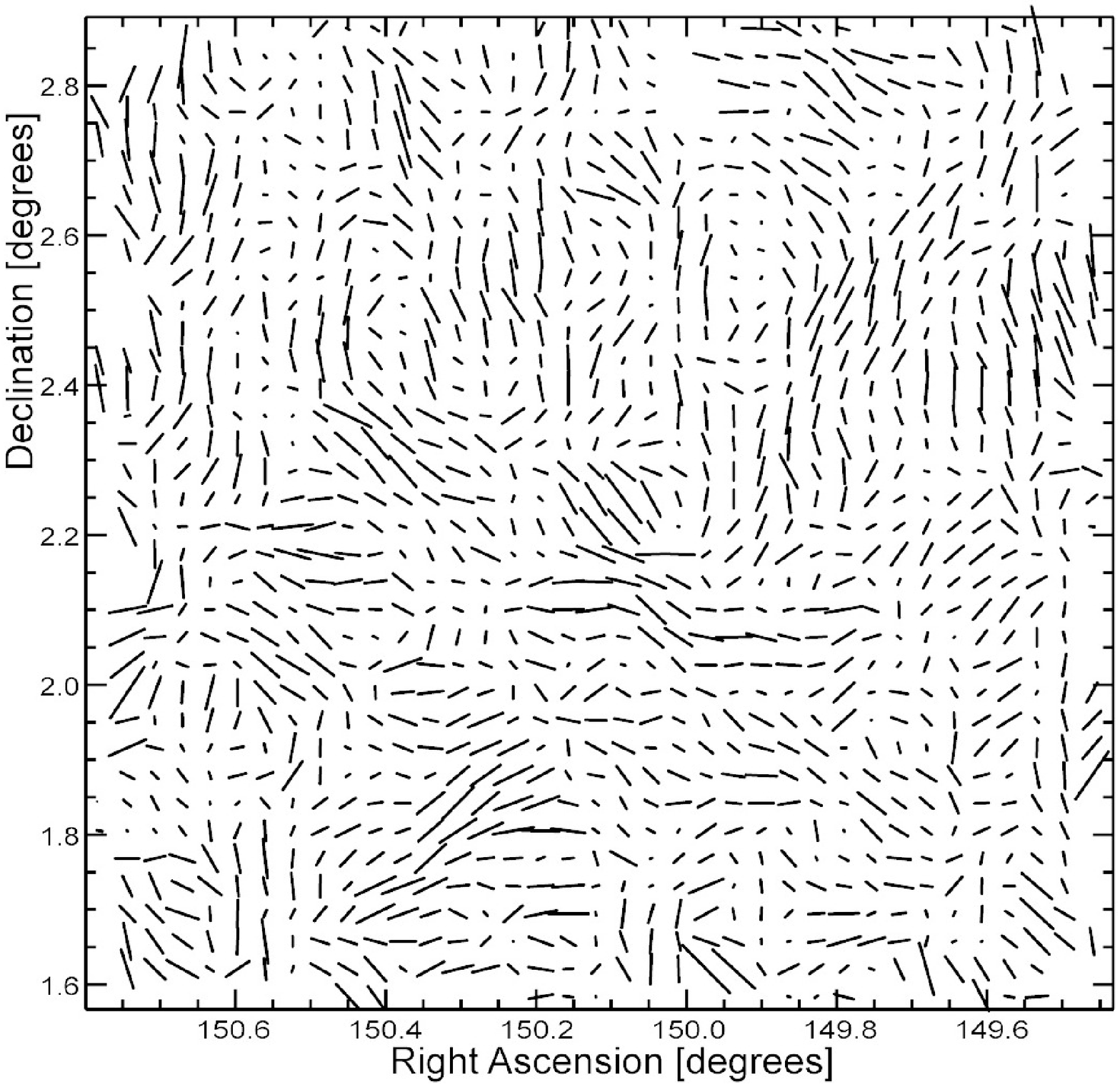}~~~~
\includegraphics[scale = 0.2]{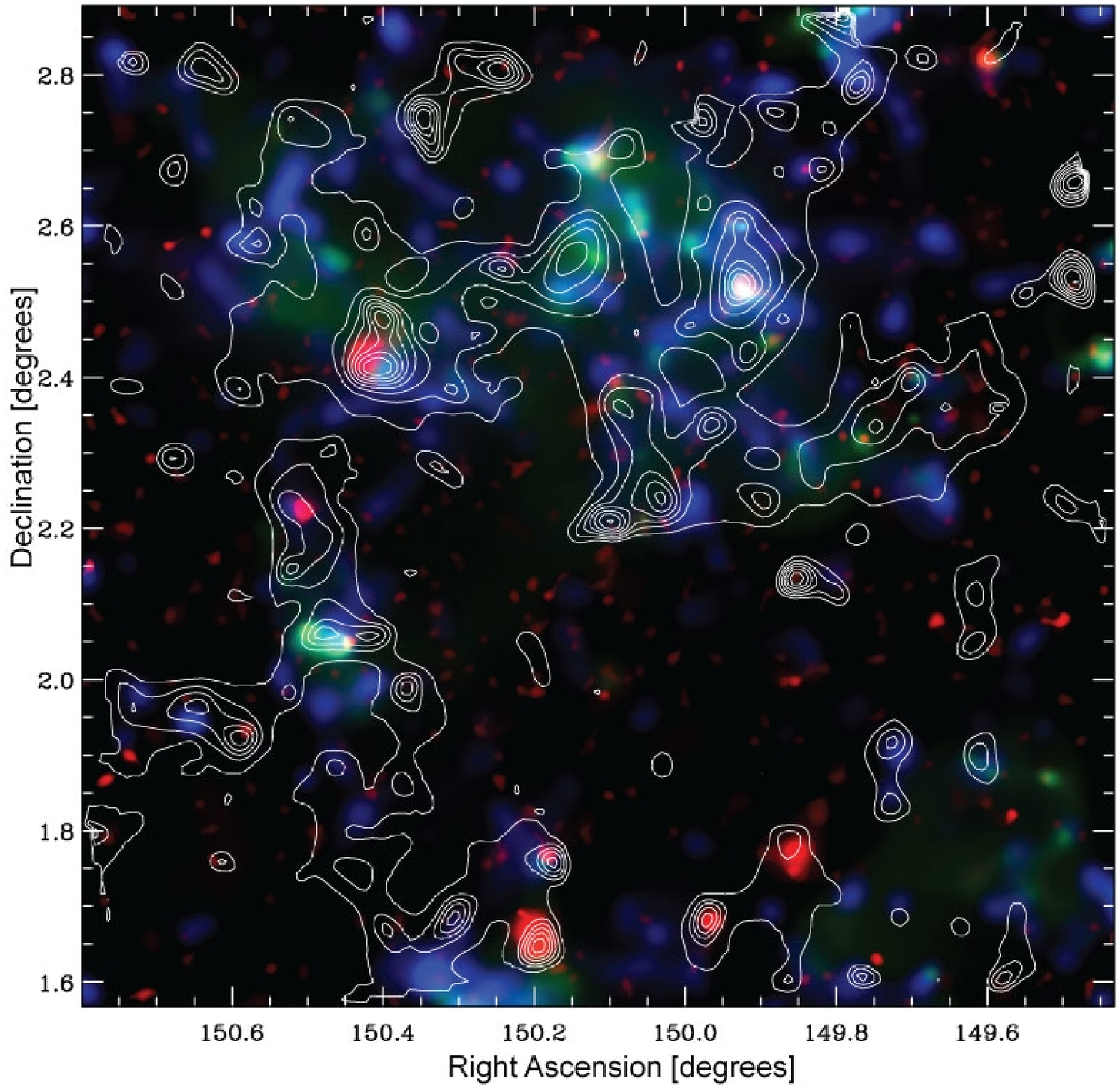}
\end{center}
\caption{{\it (Left):} The observed pattern of coherent ellipticity in background galaxies within the 2
square degree HST COSMOS survey. Each tick mark
represents the mean ellipticity of several hundred galaxies; the final analysis  uses smaller
bins containing about 80 galaxies and improves spatial resolution but adds noise. A dot in
this plot represents a circular mean galaxy; lines represent elliptical mean galaxies, with
the length of the line proportional to the ellipticity, and in the direction of the major axis.
The longest lines represent an ellipticity of about 0.06. Several circular
patterns are evident in this figure, eg (149.9, 2.5), 
and the $B-$mode signal is consistent with zero.
{\it (Right):} The reconstructed large-scale distribution of mass in front of the galaxies. 
Contours show the total distribution of mass, projected onto the plane of the
sky. Like an ordinary optical lens, gravitational lensing is most sensitive to structures
half-way between the source and the observer. The reconstruction is thus most sensitive to
mass at redshift $z=0.7$ and, to a lesser extent, all mass between redshifts 0.3 to 1.0. The
various background colours depict different tracers of baryonic light.  Green shows the
density of optically-selected galaxies and blue shows those galaxies, weighted by their
stellar mass from fits to  their spectral energy distributions. These have both been weighted
by the same sensitivity function in redshift as that inherent in  the lensing analysis. Red
shows X-ray emission from hot gas in extended sources, with most point sources removed. This
is stronger from nearby sources, but weaker from the more distant ones. (Figure credit:
Nature/R. Massey).}\label{fig3}
\end{figure}

Figure~\ref{fig3} shows the observed ellipticities of half a million distant galaxies in the
COSMOS field, deconvolved from the PSF and corrected for CTE trailing. Several
regions exhibit the circular patterns discussed above. Radial patterns are also present
on larger scales, but are much less pronounced because the density contrast in a void is
limited by the inability of density to be less than zero. A filter to detect the
patterns shown in figure 2 has been run over this data, and the measured $E$-mode signal
is presented in figure 4. Contours show the reconstructed distribution of foreground
mass: its filamentary structure is apparent. Interesting features include not only the
concentrated mass peaks at the vertices of filaments, but also the completely empty
voids. The coloured background depicts various tracers of baryons. The general correlation
between baryonic mass and total mass is striking. This is natural if the dark matter 
particles interact via gravity. All particles end in the same place via their mutual gravitational 
attraction; indeed, since the dark matter begins to forms structures first, it acts as
a scaffolding within which baryonic material is assembled.

\subsection{Galaxy clusters}

A similar picture is obtained from reconstructions of the mass in individual galaxy
clusters, which are all enveloped by a halo of dark matter. However, subtle differences between
the detailed distribution of baryonic and dark matter reflect their different
interaction properties. For example, the inner core profile of the total indicates
whether its construction was cominated by a single major merger, or gradual accretion of
smaller subhalos. The particularly flat central concentration of mass observed in
clusters could be explained by either a low level of (self-)interaction of dark matter
particles via the weak force, or by partial free-streaming of `warm' dark matter away from
the gravitational potential well. However, this picture is complicated by astrophysical
processes and events within the cluster, such as feedback, where the highly energetic
explosions of massive stars purge material from the cluster core. Even dark matter
can be affected by this, via the dynamical gravitational influence of ejected baryonic
material.

%
%

\begin{figure}[t]
\begin{center}
\includegraphics[scale = 0.3]{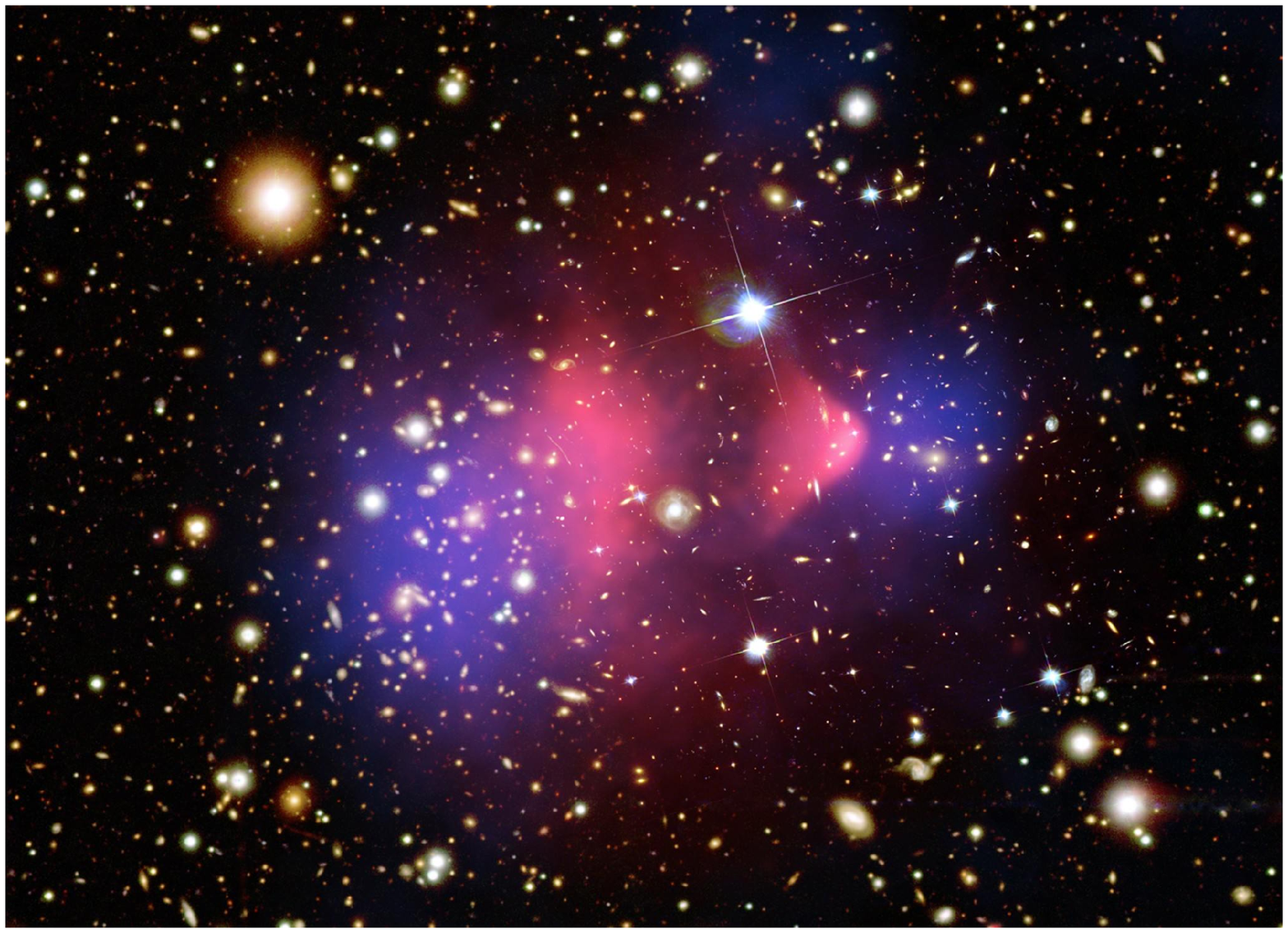}
\includegraphics[scale = 0.3]{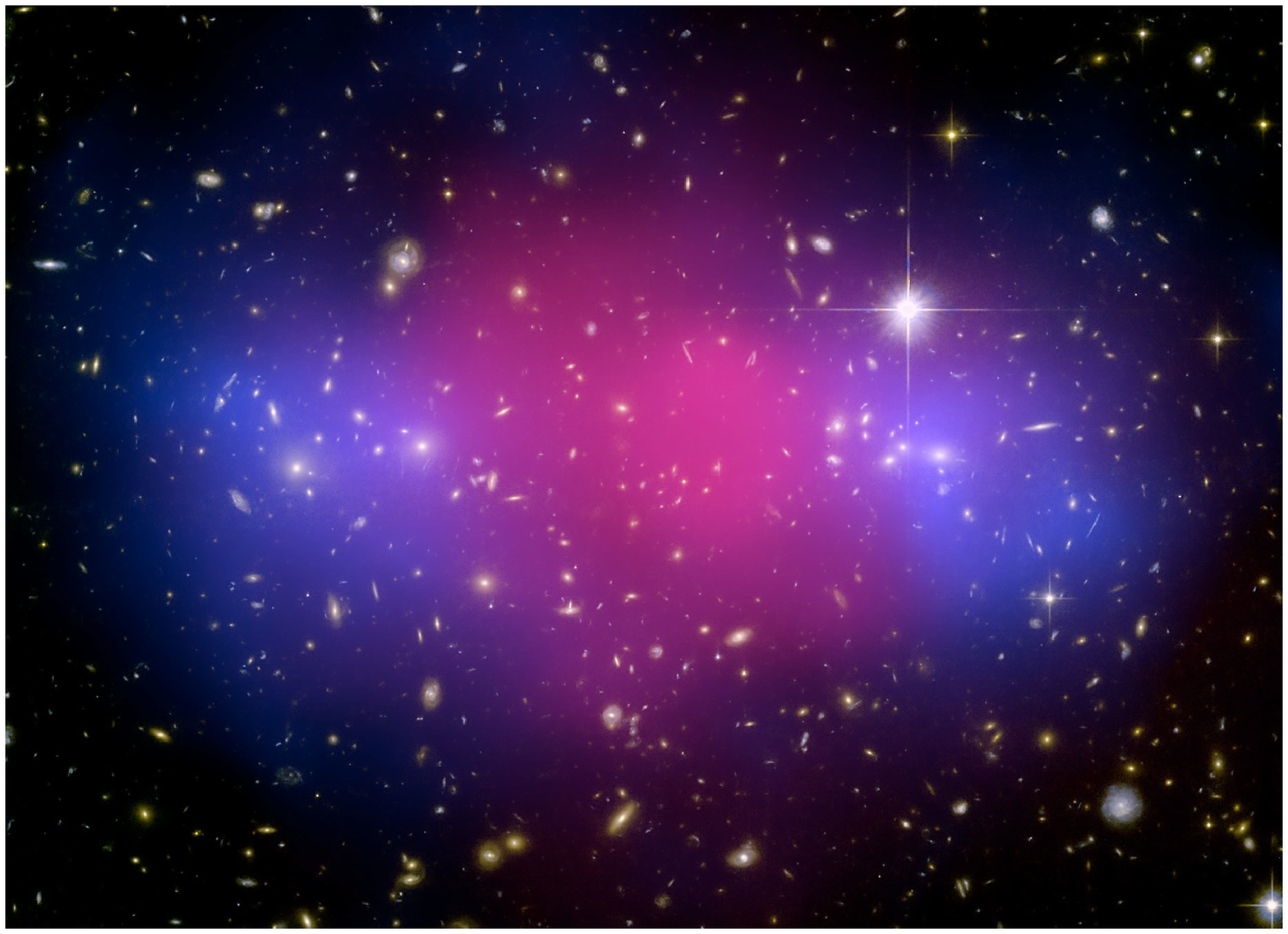}
\end{center}
\caption{
The ``bullet cluster'' 1E0657-56 and ``baby bullet'' MACSJ0025.4-1222. 
The background images show the location of galaxies, with most
of the larger yellow  galaxies associated with one of the clusters. The overlaid pink features
show X-ray emission from hot, intra-cluster gas. Galaxies and gas are baryonic
material. The overlaid blue shows a reconstruction of the total mass from measurements of 
gravitational lensing. This appers conincident with the locations of the galaxies, implying it
has a similarly small interaction corss-section. However, there is far more mass that that
present in the stars within those galaxies, providing strong  evidence for the existence
of an additional reserve of exotic dark matter. (Figure credit: X-ray: NASA/CXC/CfA/
M.Markevitch et al.; Lensing Map: NASA/STScI; ESO WFI; Magellan/U.Arizona/ D.Clowe et al.\ 
Optical image: NASA/STScI;  Magellan/U.Arizona/D.Clowe et al.; Right: NASA/ESA/M.\ Bradac et al.). }\label{fig5}
\end{figure}

The most interesting of all observed galaxy clusters is without doubt the bullet cluster
1E~0657-56, shown in figure~\ref{fig5}. The unusual separation of its constituents provides the
most direct evidence that dark matter and baryonic matter have very different properties. The
bullet cluster is strictly two clusters that collided recently (about 150 million years ago).
Rather like the scattered ejections from a particle collider, the trajectories of the detritus
allow us to determine the properties of its initial ingredients. Individual galaxies within the
clusters were well-spaced and had a very low collisional cross-section: most continued moving
during the collision, and today lie far from the point of impact. On the other hand, hot
intra-cluster gas was uniformly spread throughout the incident clusters. This had a large
interaction cross-section and -- like a cosmic car crash -- was slowed dramatically by the
collision. The two concentrations of gas, seen in X-ray emission, have now passed through each
other, but have not travelled far from the point of impact. Interestingly, the collision speed and
gas density were sufficient for a shock front to be observed in the gas from the smaller of the
two clusters, allowing the determination of the collision speed\footnote{One interesting controversy surrounding the bullet cluster has recently been resolved. The X-ray
shock front at first suggested a collisional speed at the point of impact that would be a $5\sigma$ outlier
in the range of speeds expected for such events: an unusually high figure to explain via
individual peculiar motions. However, more recent calculations took into account the acceleration
of the smaller cluster towards the larger and the relative motion of the gas clouds in the rest
frame of the smaller cluster. With these calculations, the apparent velocity implied by the shock
front is not unusal but merely high. Furthermore, the shock front that led to the discovery of this
object would not have been seen for a low-speed collision.}.

Crucially, gravitational lensing observations require an additional ingredient in the bullet
cluster. A great deal of extra mass ($\sim30-40\times$ that seen in the galaxies' stars) is
located near the galaxies, and $8\sigma$ away from the gas peaks. This mass clearly had a very low
or zero collisional cross-section, but exhibits a usual gravitational influence. Indeed, it
behaves exactly like predictions of dark matter. Measurements of its interaction cross-section,
incorporating the observed collisional speed, rule out even minimally self-interacting dark matter
models that could have explanained the flat mass profiles in cluster cores.

The visible separation between the three ingredients of each cluster is
temporary. Within another billion years, the mutual gravitational attraction of
the galaxies, gas and dark matter will have pulled them back together. They will spiral ever
closer together until they resume the usual configuration of baryonic material within a larger
dark matter cocoon.

\subsection{Galaxy groups and individual galaxies}

The gravitational lensing signal is weaker around the less massive dark matter haloes around
individual or small groups or galaxies. However, stacking the signal from many haloes in the large
SDSS and COSMOS surveys has consistently revealed a consistent central density peak of baryonic
mass, inside a halo of dark matter that exhibits several distinct components. Haloes are most
pronounced around red galaxies with older stellar populations but, for a given morphological
type, appear to have changed very little since redshift $z\sim 0.8$.

\section{Future opportunities}

\subsection{Interesting theoretical challenges}

To draw useful conclusions, or constraints on comological parameters, weak lensing measurements
must be compared to either the distribution of baryonic material or to theoretical predictions.
Weak lensing measurements are made on small scales, where the signal is strongest because mass
clumps and density grows non-linearly. However, these scales are also the most difficult to model
in $n$-body simulations because of `gastrophysical' processes; and simulations featuring modified
theories of gravity are barely possible. Observations are approaching the current precision of
theoretical models and, if future uncertainty is not to be dominated by the inability of theories
to predict the signal, advances in computational cosmological modelling must proceed apace. 

A significant image analysis challenge is posed by the  precision required in the measurement of
faint galaxy shapes -- even from high resolution, space-based images with a small PSF. A
world-wide collaboration of the weak lensing community has recently tested the performance of
existing methods in the Shear TEsting Programme (STEP). Simulated astronomical images containing
an artificial weak gravitational lensing signal were analysed blindly, such that entrants did not
know the input signal when they ran their methods, but could compare to it later. Sufficient
precision was demonstrated for current surveys. However, significant advances will be required as
larger surveys are obtained, with potentially smaller statistical errors. Improvements are
ongoing. To recruit expertise from statistical inference and computational learning fields,
STEP has now evolved into the GRavitational lEnsing Accuracy Testing 2008 (GREAT08) and 2010
(GREAT10) challenges.

Insufficient work has focused on correcting measurements of weak gravitational lensing for
the Charge Transfer Inefficiency of CCD detectors damaged by harsh radiation in orbit.
Empirical calibrations have attempted to quantify the
spurious changes in photometry, astrometry and shape measurement that are induced by the
trailing of charge. Some pipelines have also been developed to take an imperfect,
observed image then move charge back to where it belongs, pixel by pixel. This is the
ideal approach, and should be the first step in an ideal data reduction pipeline since
the trailing is created during CCD readout, the last process to happen on chip. However,
current algorithms have concentrated on long trails from species of charge trap that
affect photometry more than morphology. It remains unclear how well such techniques can be
applied to correct short trails that alter an object's shape.

\subsection{Construction of dedicated missions}

To limit the impact of the PSF on distant galaxy shapes, weak lensing needs to be observed from
above the Earth's atmosphere. Initial surveys have developed into a mature field  with the Hubble
Space Telescope. However, the current state of weak lensing measurements is rather like that of
the CMB before WMAP: a series of measurements all roughly consistent, but showing only a hint of
the high precision cosmology that could be possible. HST (and even JWST) is limited by its small
field of view. The HST COSMOS mosaic of six hundred adjacent images maps one representative region
of the sky, but is merely a proof of concept. Several efforts are proceeding in earnest to
construct a wide-field imager on a high altitude balloon (HALO) or in space (Euclid/JDEM/IDECS).
Designed for thermal stability and resistant to radiation damage, these will map the entire sky at
high precision, revealing our location in the much larger cosmic web, and directly track the
growth of structure in our expanding universe.

\section{Conclusions}

Weak gravitational lensing has long been postulated as a unique way to probe the large-scale
gravitational fields of the universe, and serious attempts to detect it have been made since the
1980s. However, it was only with the advent of large-format CCD cameras around 2000 that the tiny
signal was first detected. Indeed, the detection was so dependent upon technological innovation
that four groups independently reported detections in that one year. Further recent progress in
wide-field camera technology has rapidly advanced the field: in only nine years, weak
gravitational lensing has become an accepted staple of cosmological tests, with detailed plans for
dedicated balloon and space missions. It represents a unique way to probe the distribution of mass
in the universe, free of the usual reliance upon (often biased) tracers of electromagnetic
radiation. Interesting challenges remain to be solved in theoretical computation, practical image
analysis, and engineering design. Yet weak lensing appears to have a bright future as a mature
astrophysical and cosmological tool.

\vspace{-2mm}
\section*{References}

Bradac M., et al., ApJ 652, 937 (2006)\\
Bridle S. et al., AOAS submitted, arxiv:0802.12.14 (2008)\\
Clowe, D. et al., ApJL, 648, 109 (2006)\\
Heymans C. et al., MNRAS 368, 1323 (2006)\\
Johnston D.\ et al., ApJ submitted, arXiv:0709.1159 (2007)\\
Massey R. et al., MNRAS 376, 13 (2007)\\
Massey R. et al., ApJS, 172, 239 (2007)\\
Massey R. et al., Nature, 445, 286 (2007)\\
Refregier A., ARA\&A, 41, 645 (2004) \\
Rhodes J., et al., ApJS, 172, 203 (2007)\\
Sand D., Treu T., Ellis R., Smith G. \& Kneib, J.-P. ApJ, 674, 711 (2008)\\
Scoville N., et al., ApJS, 172, 38 (2007)\\
Schneider, P. 2005, Gravitational Lensing,
Springer-Verlag, Berlin, 273\\




\end{document}